# Experimental Study of Collective Pedestrian Dynamics

**Cécile Appert-Rolland[1], Julien Pettré[2], Anne-Hélène Olivier[3], William Warren[4], Aymeric Duigou-Majumdar[1], Etienne Pinsard[5], Alexandre Nicolas[5]**
[1] Laboratory of Theoretical Physics
CNRS (UMR8627), Univ. Paris-Sud, Univ. Paris-Saclay, F-91405 Orsay, France
Cecile.Appert-Rolland@th.u-psud.fr
[2] Inria, Univ Rennes, CNRS, IRISA - UMR 6074, F-35000 Rennes, France
[3] Univ Rennes, Inria, M2S - EA 7470, F-35000 Rennes, France
[4] CLPS, Brown University – Providence, RI, USA
[5] Laboratoire de Physique Théorique et Modèles Statistiques
CNRS (UMR8626), Univ. Paris-Sud, Univ. Paris-Saclay, F-91405 Orsay, France

***Abstract*** - We report on two series of experiments, conducted in the frame of two different collaborations designed to study how pedestrians adapt their trajectories and velocities in groups or crowds. Strong emphasis is put on the motivations for the chosen protocols and the experimental implementation. The first series deals with pattern formation, interactions between pedestrians, and decision-making in pedestrian groups at low to medium densities. In particular, we show how pedestrians adapt their headways in single-file motion depending on the (prescribed) leader's velocity. The second series of experiments focuses on static crowds at higher densities, a situation that can be critical in real life and in which the pedestrians' choices of motion are strongly constrained sterically. More precisely, we study the crowd's response to its crossing by a pedestrian or a cylindrical obstacle of 74cm in diameter. In the latter case, for a moderately dense crowd, we observe displacements that quickly decay with the minimal distance to the obstacle, over a lengthscale of the order of the meter.

## 1. Introduction

The motion of a particle in the midst of a crowd often differs dramatically from its motion in free space [1]. This is particularly true for pedestrians, who move in one or two-dimensional space, where trajectories are more likely to intersect than in three dimensions. The conflicts generated by space-sharing do not arise only because of physical constraints (which preclude overlaps, in particular), but also because of individual preferences, which render some situations unfavorable, in spite of being physically possible, for instance coming in close contact with another pedestrian or markedly deviating from one's target direction. When such conflicts arise, pedestrians effectively interact, notably to avoid collisions, in converging or intersecting flows; they also interact when a collective decision must be made. These interactions are complex and there is no ground rule to predict them. Besides, given their non-additivity, many-body interactions cannot be straightforwardly derived from simpler situations, such as encounters between two pedestrians.

Therefore, reference (empirical) data are particularly needed regarding the way pedestrians adapt their motion in crowds, in order to test models for pedestrian dynamics [3] or to assess its impact on critical infrastructure [2]. Indeed, too often, these models are calibrated using only 'macroscopic' observables. To collect reference data, two competing approaches exist: passive observations of real situations (streets, shopping malls, railway stations, etc.) and controlled experiments with voluntary participants. The former approach captures the complexity of real crowds, but involves many uncontrolled parameters which complicate the interpretation of the observations. Prepared experiments, on the other hand, render the tracking of pedestrians easier, with a variety of possible devices, and allow one to control parameters such as the density and the geometry of the environment, and to repeat the same protocol several times in order to make a statistical analysis possible. On the downside, the participants' behaviors and the composition of the crowd may be biased by the experimental conditions. Rather than competing possibilities, the two approaches thus turn out to be complementary.

In this paper, we report on two series of controlled experiments, conducted in the frame of two different collaborations and designed to study how pedestrians adapt their trajectories and velocities in groups or crowds. We will emphasize the motivations for the experimental protocols and their implementation, while deferring an exhaustive analysis of the results to a future publication.

The first series of experiments (PEDINTERACT project) considers situations of low to intermediate densities, involving a lot of conflicting trajectories or some collective decision process. The second series (PERCE-FOULE project) focuses on crowds at high density and likens them to a deformable medium.

## 2. Scanning the interactions between pedestrians at low or intermediate densities

Even at low or intermediate densities, pedestrians must interact to coordinate their motion. First comes, of course, the constraint that the chosen trajectories do not lead to collisions. Then, in some situations, collective decisions need to be made about where to head for or what route to choose. The experiments conducted in the frame of the PEDINTERACT project are aimed at better understanding the many-body interactions between pedestrians involved in these situations. We designed several protocols, some focusing on the need to solve multiple threads of collisions, while others put the focus on collective decisions.

### 2.1. Description of the PEDINTERACT experiments
The collaboration involves J. Pettré (INRIA Rennes), A.-H. Olivier (M2S Rennes), W. Warren (CLPS, Brown Univ.), C. Appert-Rolland (LPT – Orsay), and their students. The experiments took place at the Immermove platform of the M2S laboratory in Rennes (France) on $30^{th} – 31^{st}$, March 2016, in a sport hall equipped with 26 VICON cameras. On each of the two days, 36 to 38 adult volunteers took part in the experiment. They were wearing a bike helmet carrying 4 markers detectable by the VICON motion capture system and arranged in a unique configuration. For each experimental protocol, several realizations were performed in order to allow for a statistical analysis. A post-processing stage allows to reconstruct the individual trajectories.

### 2.2. Protocols
***Swarm behavior:*** At the start, participants (unaware of the purpose of the experiment) were confined in a 8x8 $m^2$ area. They were assigned initial heading directions making an angle of 90, 180, 270, or 360 degrees. They were then asked to walk around in the sport hall. The purpose of the experiment was to observe whether herding or correlated collective motion spontaneously emerges.

***Group intersections:*** Two groups of 20 people each were asked to "walk straight across the room and pass through another group of participants". The groups were not constrained laterally: they could split or become porous and merge after the intersection. The initial density within each group (around 1 ped/$m^2$ or 0.45 ped/$m^2$) and the angle at which the groups intersect (0, 15, 30, 90, 150, 165, or 180 degrees) were varied. For each set of parameters 12 realizations were performed, during which the composition of the groups was changed 4 times. Six additional realizations of the 90 degrees crossings were conducted with 4 pillars delimiting the intersection area, to mimic the junction between two perpendicular corridors, in which case the formation of diagonal stripes was predicted and observed [4].

***Gate passing:*** A gate of width 80cm or 240 cm was crossed either unidirectional by a group of 40 people or bidirectionally by 2 groups of 20 people. Besides, the influence of the initial crowd density on the heading towards the gate was studied using an even larger gate (800cm). Each situation was repeated 6 times.

***Collective turning decision:*** A group of 40 people, initially packed with a density of 1 or 0.45 ped/$m^2$, was asked to "walk across the room, and make a left or right turn, while staying together as a group". The instruction to turn right or left was given to the group before each realization. As there was no official group leader, a non-verbal decision had to be reached about when to turn. Sometimes a single leader emerged, but more often a leading small subgroup took charge. An intriguing question is whether there exists a critical size for this leading subgroup above which others will automatically follow, and how followers will deal with conflicting information if two leading subgroups go in different directions. Each case (left or right turn, initial density) was repeated 5 times.

In the next section, we will present into deeper details one last protocol involving collective and individual decisions, as well as some results.

### 2.3. Unidimensional flow without density constraint

The last protocol focuses on unidirectional flows. Under these conditions, conflicts arise whenever the pedestrians' free velocities are not exactly equal, and induce interactions. While this situation had already been studied by several groups at fixed density and with a prescribed trajectory path [5-8], we were interested in finding out how the pedestrians adapt if they are given more freedom.

For that purpose, we let the pedestrians choose their trajectories and – as a direct consequence – determine the density collectively. To do so, we asked a fixed number of pedestrians (18) to walk along a circle, without further specification: Neither a center nor a radius were prescribed. Thus, a decision had to be made about the formation of the circle, which resulted both from a following behavior (interaction with the predecessor) and from a more global perception of the collective trajectory (see Fig. 1).

An intermediate case was also considered, in which pedestrians had to walk along a straight line across the sport facility. In this case, the path was fixed and the leader was imposing a maximal velocity, either his free velocity or a prescribed low velocity. Meanwhile, followers were free to choose their headways, or, in other words, the local density. Table 1 provides a summary of the realizations.

Table 1: Realizations of unidirectional pedestrian flows with unconstrained density.

| Trajectories | Number of repetitions | Number of pedestrians |
|---|---|---|
| Circle | 6 | 18 |
| Straight with free velocity | 7 | 36 |
| Straight with low velocity | 3 | 36 |

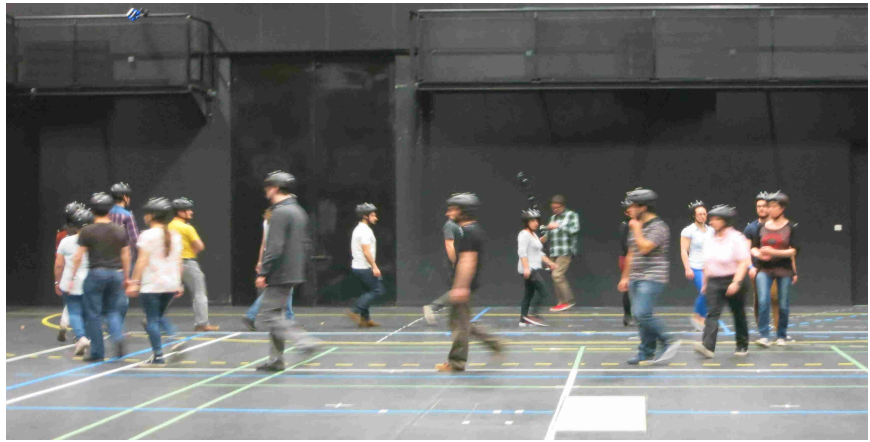

Fig. 1: Snapshot taken during a unidirectional flow experiment, in which the participants walked along a circle that they were free to choose.

***Fundamental diagram:*** For each pedestrian, we define the instantaneous density as the inverse distance between the middle position of the pedestrian and its predecessor. In the case of a circular trajectory, the distance is the arc length. Instantaneous velocities are evaluated after filtering out the steps. To do so, we use a second-order low pass butterworth filter with a cut-off frequency of 0.8Hz.

Computing these variables gives access to fundamental diagrams, an example of which is shown on Fig. 2-left for two realizations of straight walks. In one of them, the leader of the line walked at his free velocity, while on the other he was told to walk slowly. Recalling that followers are free to choose their headways, we notice that they chose it such that the resulting mean density-velocity point (black circle and diamond on Fig. 2-left) follows the fundamental diagram observed in similar experiments conducted at fixed density (Fig. 2-right) [5]. The relation between velocity and density thus seems to be similar if one imposes the density or the velocity.

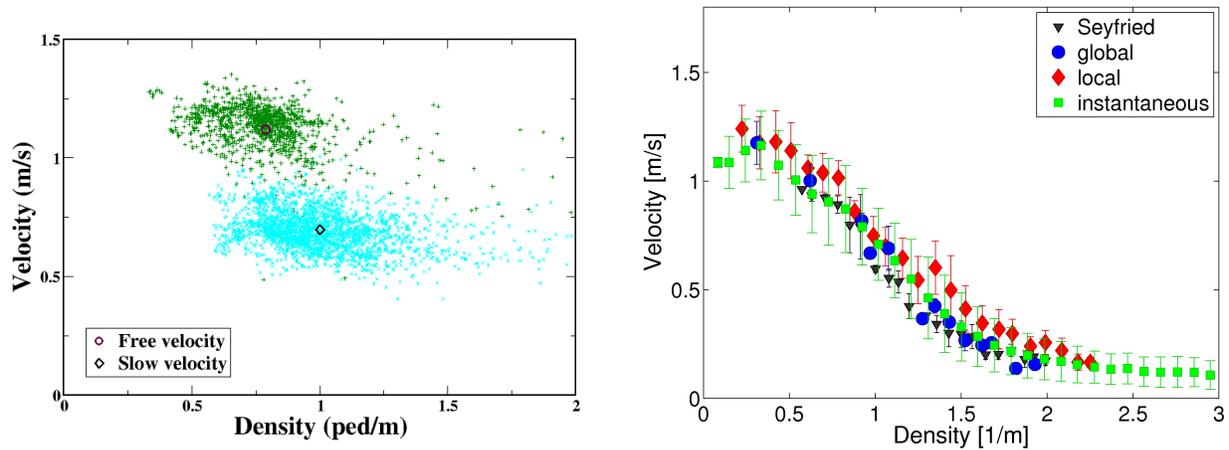

Fig. 2: Fundamental diagrams *(right)* for a circular path with density constraint (from [5]) and *(left)* for two realizations of straight unidirectional flows, with a leader walking with free *(green)* or slow *(blue)* velocity; Small plus signs correspond to instantaneous (velocity, density) values, while thick symbols *(circle and diamond)* give the average values.

### 3. Response of dense crowds to local perturbations

We now turn to a study of static crowds at higher densities. In real life, these conditions are encountered in metro trains, at mass gatherings, or in front of a facility or a shop just before its opening (for instance, on Black Fridays in the US). Insight into the collective response of these crowds would be particularly valuable, because current agent-based models are generally not calibrated for these dense conditions and their predictions can therefore not be relied on.

Because of the density, pedestrian motion is strongly constrained sterically and physical contacts play a more and more important role in their interactions. It is appealing to handle the crowd as a dense physical medium and, effectively, research has pointed to quantitative similarities between the dynamics of dense crowds and granular media, for instance [9, 10] . However, one should not forget that decision-making processes are also present and an exciting question is how the latter couple with the strong mechanical constraints.

Experimentally, the study of very dense crowds raises some challenges in terms of safety (in controlled experiments) and detection of tightly packed pedestrians. Overcoming these challenges, we performed controlled experiments to characterize the response of dense crowds of standing people to the crossing of an intruder, within the frame of Project PERCE-FOULE (collaboration between A. Nicolas at LPTMS and C. Appert-Rolland at LPT). The intruder will either be a real pedestrian making through the crowd or it will be abstracted into a moving cylindrical obstacle. Quite interestingly, these two experimental cases echo recent similar studies performed numerically on a minimalistic pedestrian model [11] and experimentally on a granular medium penetrated by an intruder [12,13], respectively. We aim to describe the local reorganization of the crowd triggered by these intrusions.

### 3.1. Description of the PERCE-FOULE experiments

After their formal approval by the local ethics committee (C3E), the experiments took place in in a sport facility at University Paris-Sud in Orsay (France) on 26th, June 2017. They involved up to 39 voluntary participants, mostly students aged 20-30, with around 20% of women.

These participants were asked to stand in a delimited area (typically 4mx2.5m) and behave as if they were on an underground platform. They were wearing hats, with home-made bar codes (these are not used in the following), to facilitate the tracking. They were filmed from above by two TomTom Bandit cameras (60 frames per second, 1080p) and one Microsoft Kinect sensor (which provides the field of distances between objects and the focal plane of the sensor), hanging at an altitude of around 4 meters.

As already mentioned, two types of crossing experiments were performed, namely (i) pedestrian crossings, in which one participant was asked to walk to the opposite end of the crowd, (ii) crossing of a cylindrical obstacle. The role of the obstacle was played by one of the organizers, standing in a cylinder made of hula hoops of 74 cm in diameter and moving along a straight line. Type (ii) was expected to trigger larger deformations, with well-characterized obstacle shape and motion. Prior to each series of crossings, the crowd density was adjusted to an intermediate (~2 ped/m²), high (~4 ped/m²) or very high (~6 ped/m²) value by controlling the size of the standing zone while the pedestrians' orientation was prescribed to be either random, or facing the arrival of the obstacle. For each experimental condition, we performed 6 to 10 repetitions. Snapshots of one crossing by an obstacle are shown in Fig. 3.

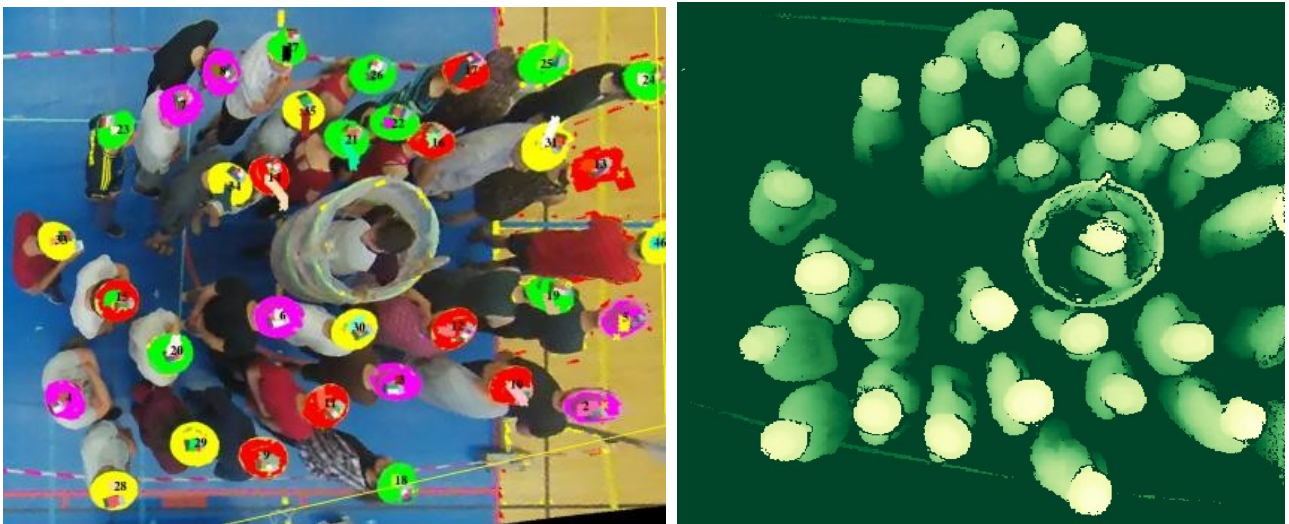

Fig. 3: Snapshot of an experiment consisting in the crossing of a moderately dense crowd by a cylindrical obstacle. (*Left*) Image from the camera, in which the detected hats are highlighted and the tracked trajectories are shown. (*Right*) Image from the Kinect sensors. Light colors correspond to pixels that are closer to the focal plane of the camera (*i.e.*, higher).

### 3.2. Video analysis

30 images were extracted for every second of video footage (videos are thus downsampled) and corrected for the optical distortion of the lens (barrel distortion). We noticed that the cameras were rotating during the recording; this was corrected manually by applying rotations to each image so that given landmarks on the floor coincide between images.

We developed a home-made software in Python to detect and track the participants' hats on the basis of their colors. The algorithm proved very efficient and yielded almost no false negatives: virtually all hats were correctly tracked. Post-processing manual corrections allowed us to remove some spurious detections on the floor (false positives) and to track the obstacle.

The correspondence between the hat positions in pixel coordinates and the ground coordinates of the bodies was obtained by
1) finding the affine function that provides the best fit between the pixel coordinates of a subset of detected hats and the corresponding 'body' positions (defined as the middle of the feet, as determined manually when the person stands more or less straight) ;
2) converting pixel coordinates into ground coordinates via linear interpolation between reference points.

The lens distortion (once corrected), the rotation of the cameras (once corrected) and the conversion between pixel coordinates and ground coordinates were found to create an experimental uncertainty of only 5 cm or less. On the other hand, the correspondence between hats and 'bodies' (step 1 above) yielded an uncertainty of around 20 cm for not too inclined pedestrians, as we found by comparing estimated 'body' positions to manually measured positions on the videos [Fig. 4(left)]. This value is in line with our expectations, since we did not correct for the different participants' heights [14], nor for the variations in the inclinations of their chests. Figure 4(right) corroborates this upper bound on the spatial inaccuracy by comparing trajectories that were reconstructed using videos from the first camera and from the second one. As a matter of fact, in this figure, the uncertainty is slightly lower than expected (less than 15 cm).

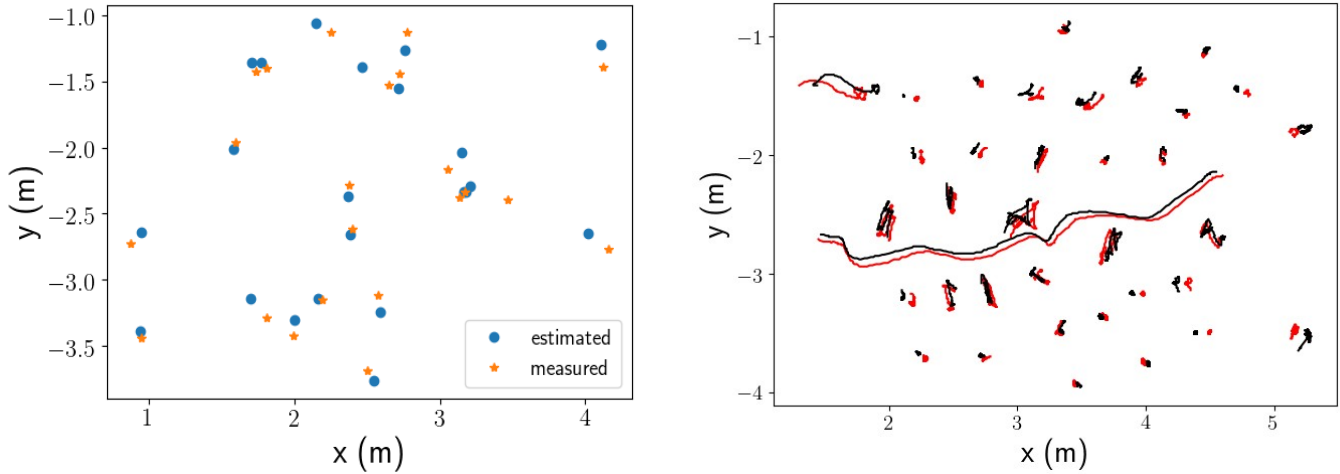

Fig. 4: (*Left*) Comparison between estimates of 'body' positions using the detected hat positions and actually measured positions. Data from different series of experiments are gathered here. (*Right*) Comparison between trajectories reconstructed using videos from the first camera (*red*) and the second one (*black*) in one crossing experiment.

### 3.3. Preliminary results

In Fig. 5 we show the traces of the participants' motion in response to the crossing of the cylindrical obstacle, for an experiment at intermediate pedestrian density. The first conspicuous observation is that larger moves are made by people who stand close to the obstacle's route, which is quite intuitive. Less than two meters away from it, pedestrians barely move owing to the crossing, at this intermediate density. Furthermore, pedestrians do not appear to anticipate the passage of the obstacle much, as far as one can judge by the evolution of the trajectories with time.

Further analysis of these results will rely on averaging over nominally identical realizations and will be enlightened by comparison with (i) the displacements of grains in a two-dimensional packing due to a moving immersed intruder [12,13], (ii) numerical simulations of passage of an individual through a crowd [11]. In the granular setup, dramatic changes were observed as the packing fraction was increased towards the jamming point, with much larger force fluctuations and the disappearance of the cavity behind the intruder, through recirculation. Such cavity was also observed behind our cylindrical obstacle and was closed in a time scale of the order of the second after its passage at high enough density (compare the middle and right panels of Fig. 5). Besides, at a density just below the jamming density, the crosswise ($x$) profile of the average instantaneous streamwise ($y$) velocity of the grains obeyed the scaling

$$u_y(x) \sim (x-x_0) e^{\frac{-(x-x_0)}{\lambda_x}},$$

with a lengthscale $\lambda_x \sim 2d$ ($d$ is the grain diameter) that increases slightly with the size of the intruder. This seems compatible with our observations in dense pedestrian crowds, but it will be interesting to test whether the latter can also be described by the above formula.

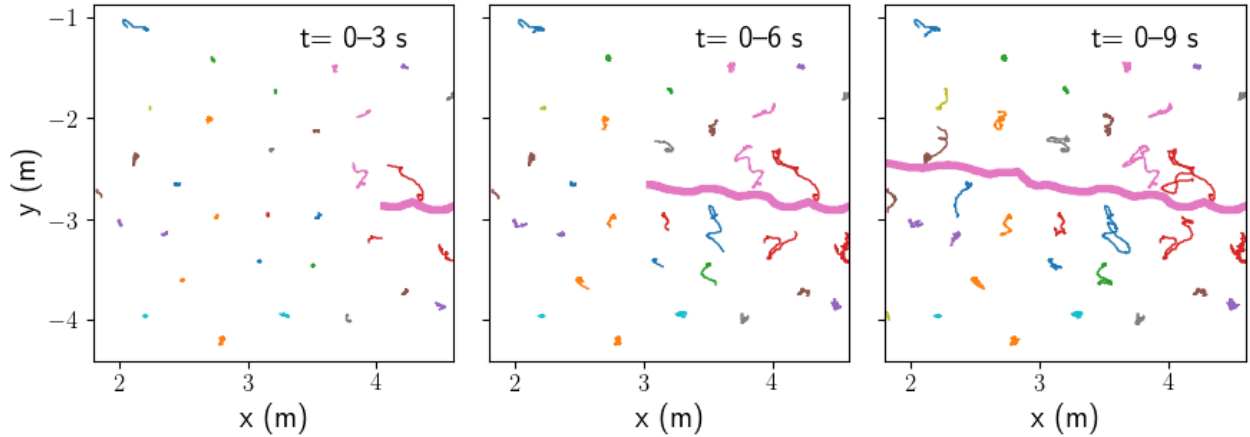

Fig. 5: Pedestrians' trajectories in the ground frame, in response to the crossing of a cylindrical obstacle, over the time windows indicated in the legends. The trajectory of the center of the cylinder is shown as a thick pink line and pedestrians were initially oriented in random directions.

## 4. Conclusion

In summary, we conducted experiments designed to shed light on the way pedestrians adapt their motion in order to cope with the constraints associated with walking or standing amid a group or a crowd. In particular, we presented results (i) on the pedestrians' adjustments of their headways in single-file motion where the leader's velocity is prescribed and (ii) on the response of a dense crowd to its crossing by a cylinder, drawing tentative parallels with the response of a two-dimensional granular packing.

We would like to conclude on a short note regarding the diverse methods for data collection presented here. VICON motion sensors provide temporally and spatially accurate localizations of the pedestrians' heads, but are a cumbersome and expensive system. Action cameras are cheaper and more portable, but are also slightly less accurate and do not directly measure the vertical coordinate (in our case leading to an experimental uncertainty on the positions close to 20cm, in part because we did not correct for the different participant's heights). Both methods require the participants to wear specific devices to facilitate their detection (although machine-learning algorithms may render this unnecessary). In contrast, depth-field sensors such as Microsoft Kinect cameras do not require pedestrians to wear specific markers or hats and are less invasive. On the downside, their field of view and spatial resolution are smaller than those of the other devices. Nevertheless, it is important to bear in mind that, besides the shortcomings of specific measurement devices, a major contributor to the uncertainty of the detected positions is inherent in the projection of the three-dimensional shapes of pedestrians onto $(x,y)$-positions. For instance, the position of the head with respect to the body is very sensitive to the (varying) inclination of the chest. In the future, to get a better grasp of the dynamics of dense crowds, which feature small spacings between pedestrians, it might become necessary to turn to a more suitable representation of the space actually occupied by the pedestrians' bodies.

## Acknowledgements

We thank all participants in the experiments and, above all, the students and researchers who took an active part in their organization: Tom Marzin, Ioannis Touloupas, Aymeric Duigou-Majumdar, and Étienne Pinsard for the PERCE-FOULE experiments; Armel Cretual, Anthony Sorel, Dany Saligaut, Trenton Wirth, Gregory Dachner for the PEDINTERACT experiments.

The PERCE-FOULE project was funded by a grant from French Labex PALM (ANR-10-LABX-0039-PALM).
PEDINTERACT experiments were funded by the Inria SIMS associated team as well as the ANR PERCOLATION ANR-13-JS02-0008 project.
PEDINTERACT Project was also funded by US National Science Foundation grant BCS-1431406